\documentclass[prx,reprint,superscriptaddress,twocolumn,showkeys]{revtex4-1}
\usepackage[utf8]{inputenc} 
\usepackage{amsmath}
\usepackage{braket}
\usepackage{graphicx}
\usepackage{pgfplots}
\pgfplotsset{compat=1.14}
\usepackage{csquotes}
\usepackage{hhline}
\usepackage{amssymb}

\usepackage{dcolumn}
\usepackage{tabularx}
\usepackage[colorlinks=true,linkcolor=blue,citecolor=blue,urlcolor=blue]{hyperref}
\usepackage{braket}
\usepackage{float}
\usepackage{listings}
\usepackage{color}
\usepackage{subcaption}
\usepackage{placeins}
\definecolor{mygreen}{rgb}{0,0.6,0}
\definecolor{mygray}{rgb}{0.5,0.5,0.5}
\definecolor{mymauve}{rgb}{0.58,0,0.82}

\newcolumntype{C}{>{\centering\arraybackslash}X}

\begin{document}

\title{Quantum Cost Efficient Scheme for Violating the Holevo Bound and Cloning in the Presence of Deutschian Closed Timelike Curves}

\author{Harshavardhan Reddy Nareddula}
\email{harsha.rn@niser.ac.in}
\affiliation{School of Physical Sciences,\\ National Institute of Science Education and Research, HBNI, Jatni 752050, Odisha, India}
\author{Bikash K. Behera}
\email{bkb13ms061@iiserkol.ac.in}
\author{Prasanta K. Panigrahi}
\email{pprasanta@iiserkol.ac.in}
\affiliation{Department of Physical Sciences,\\ Indian Institute of Science Education and Research Kolkata, Mohanpur 741246, West Bengal, India}

\begin{abstract}
Brun \emph{et al.} [Phys. Rev. Lett. \textbf{102}, 210402 (2009)] showed that in the presence of a Deutschian closed timelike curve (D-CTC), one could violate the Holevo bound. It is possible to utilize the Holevo bound violation to encode $n$-bit classical information in a single qubit. Here we demonstrate a new quantum cost efficient scheme, for storing and retrieving $n$-bit classical information faithfully in the presence of a D-CTC in violation of the Holevo bound. We also propose a new protocol for cloning a qubit in the presence of a D-CTC. In both the schemes, the quantum cost is found to be of order $O(n)$, which provides an advantage over the existing schemes having quantum cost of exponential order.
\end{abstract}
 
\begin{keywords}{Quantum Cloning, Deutschian Closed Timelike Curves, Violation of Holevo Bound}\end{keywords}

\maketitle
\section{Introduction \label{qied_Sec1}}
Closed Timelike Curves (CTC) arise as possible solutions to the Einstein's field equations \cite{qdctc_GodelRMP1949}. However, their existence is a subject of debate due to apparent paradoxes associated with time travel, such as the ``grandfather paradox". In his seminal work \cite{qdctc_DeutschPRD1991}, Deutsch proposed a model for CTCs by setting aside the details of the spacetime geometry and working with the tools of quantum information. ``Deutschian" CTCs (D-CTCs) resolve the causality paradoxes by imposing a self-consistency condition on the system traversing the CTC. Deutsch imposes that the density matrix of the CTC system is the same while entering and exiting the wormhole \cite{qdctc_MoulickSciRep2016}. Formally,
\begin{equation} \label{qdctc_Eq1}
\rho_{CTC} = Tr_{CR}(U(\rho_{CR}\otimes\rho_{CTC})U^\dagger)
\end{equation}
where $\rho_{CR}$ and $\rho_{CTC}$ are the density matrices of the chronology respecting (CR) system \cite{qdctc_DeutschPRD1991} and the closed timelike curve (CTC) system respectively. Here $U$ denotes the interaction unitary. Brun \textit{et al.} \cite{qdctc_BrunPRL2009} showed that one could perfectly distinguish non-orthogonal states in the presence of a D-CTC. They provide a recipe to construct a D-CTC assisted circuit to map $N$ distinct non-orthogonal states to the standard orthonormal basis of an $N$-dimensional space. A consequence of this is the violation of Holevo bound \cite{qdctc_GoswamiarXiv2017}. Thus, it is possible to encode $n$-bit classical information in a single qubit and retrieve it faithfully. If we were to follow Brun's recipe, however, we would be required to construct $2^n$ number of $2n$-qubit gates. Here we present a quantum cost efficient scheme for retrieving $n$-bit classical information encoded in a specific manner in a single qubit. The ability to distinguish non-orthogonal states also leads to the violation of the no-cloning theorem. Several schemes were presented to clone a quantum system \cite{qdctc_AhnPRA2013,qdctc_BrunPRL2013}. Although an arbitrary state cannot be faithfully cloned, the fidelity of cloning increases as the number of ancillary qubits increase in these schemes. The quantum cost of these schemes is also exponential in terms of the number of ancillary qubits. We also show that our scheme can be modified to employ cloning of a qubit, while maintaining the quantum cost linear in terms of the number of ancillary qubits.

IBM quantum experience has provided five- and sixteen-qubit quantum computers to perform quantum experiments around the world via cloud based services. For the past one year, a number of quantum computational tasks have been performed using this platform. Some of them include testing of quantum algorithms \cite{qdctc_GangopadhyayQIP2018,qdctc_ColesarXiv:1804.03719}, quantum simulation \cite{qdctc_LiertaarXiv2018,qdctc_ZhukovQIP2018,qdctc_KapilarXiv:1807.00521,qdctc_ViyuelanpjQI2018,qdctc_HegadearXiv:1712.07326}, quantum artificial intelligence \cite{qdctc_RodriguezarXiv:1711.09442}, quantum machine learning \cite{qdctc_ZhaoarXiv:1806.11463}, quantum algorithms for hard problems \cite{qdctc_SrinivasanarXiv2018,qdctc_DasharXiv2018}, algorithms for quantum games \cite{qdctc_PalRG.2.2.19777.86885,qdctc_MahantiRG.2.2.28795.62241}, realization of quantum devices \cite{qdctc_BeheraarXiv2017,qdctc_BeheraarXiv2018,qdctc_BeheraQIP2017}, quantum information sciences \cite{qdctc_FedortchenkoarXiv:1607.02398,qdctc_SisodiaQIP2018,qdctc_VishnuQIP2018,qdctc_MajumderarXiv:1803.06311,qdctc_SatyajitQIP2018,qdctc_SisodiaPLA2017,qdctc_KalraarXiv2017,qdctc_RoyarXiv2017}, quantum error correction algorithms  \cite{qdctc_SingharXiv:1807.02883,qdctc_HarperarXiv:1806.02359,qdctc_WillscharXiv:1805.05227,qdctc_GhoshQIP2018}, quantum circuit optimization techniques \cite{qdctc_ZhangarXiv:1807.01703,qdctc_DueckarXiv2018}. 

\section{Encoding an $n$-bit register in a single qubit \label{qied_Sec2}}
Suppose we are required to encode an $n$-bit register $(a_n...a_1)$ pertaining to the numerical value $k$ in a single qubit, where $k$ can assume values ranging from 0 to $2^n-1$. Then we need to map the possible values of $k$ to a set of $2^n$ distinct non-orthogonal states of the qubit. In our scheme, we choose a set of $2^n$ evenly spaced pure states lying on the XZ-plane of the Bloch sphere for this purpose. Explicitly, we prepare the encoded qubit in the state $\Ket{\psi_k}$ to represent $k$ where 
\begin{equation}\label{qdctc_psik}
    \Ket{\psi_k} = \cos{\frac{\pi k}{2^n}}\Ket{0} + \sin{\frac{\pi k}{2^n}}\Ket{1}
\end{equation}
\begin{figure}[]
\includegraphics[width=\linewidth]{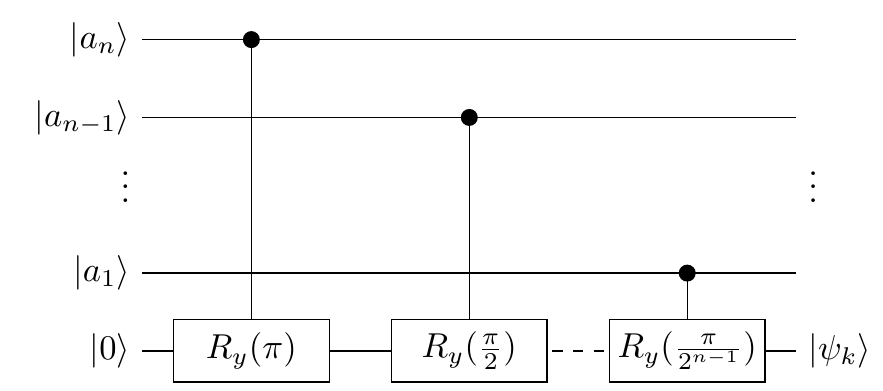}
\caption{\textbf{Encoding of an $n$-bit register in a single qubit.} An encoding qubit is initialized to state $|0\rangle$. $\Ket{\psi_k} = \cos{\frac{\pi k}{2^n}}\Ket{0} + \sin{\frac{\pi k}{2^n}}\Ket{1}$ is prepared, which encodes the $n$-bit register $(a_n...a_1)$ corresponding to value $k$, after the application of $n$ controlled rotation operations on the encoding qubit.} 
\label{qdctc_Fig1}
\end{figure}
Note that a (anti-clockwise) rotation of the state $\Ket{0}$ along the Y-axis by an angle of $\frac{2\pi k}{2^n}$ yields $\Ket{\psi_k}$. Such a state can be prepared by employing the quantum circuit depicted in Fig. \ref{qdctc_Fig1}. Here $R_y(\theta)$ represents the unitary operator performing an anti-clockwise rotation along the Y-axis of the Bloch sphere by an angle of $\theta$. The encoding qubit is initialized in the state $\Ket{0}$. The controlled rotations performed on the qubit add up to an overall $\frac{2\pi k}{2^n}$ rotation along the Y-axis, thus preparing $\Ket{\psi_k}$. Explicitly, the circuit performs the following transformation,
\begin{equation}
    \Ket{0} \rightarrow R_y\Big(\sum_{i=1}^n \frac{\pi a_i 2^i}{2^n}
    \Big)\Ket{0} = R_y\Big(\frac{2\pi k}{2^n}\Big)\Ket{0} =  \Ket{\psi_k}
\end{equation}

\section{Retrieving the $n$-bit register from the encoded qubit}
\begin{figure*}[]
\includegraphics[width=\linewidth]{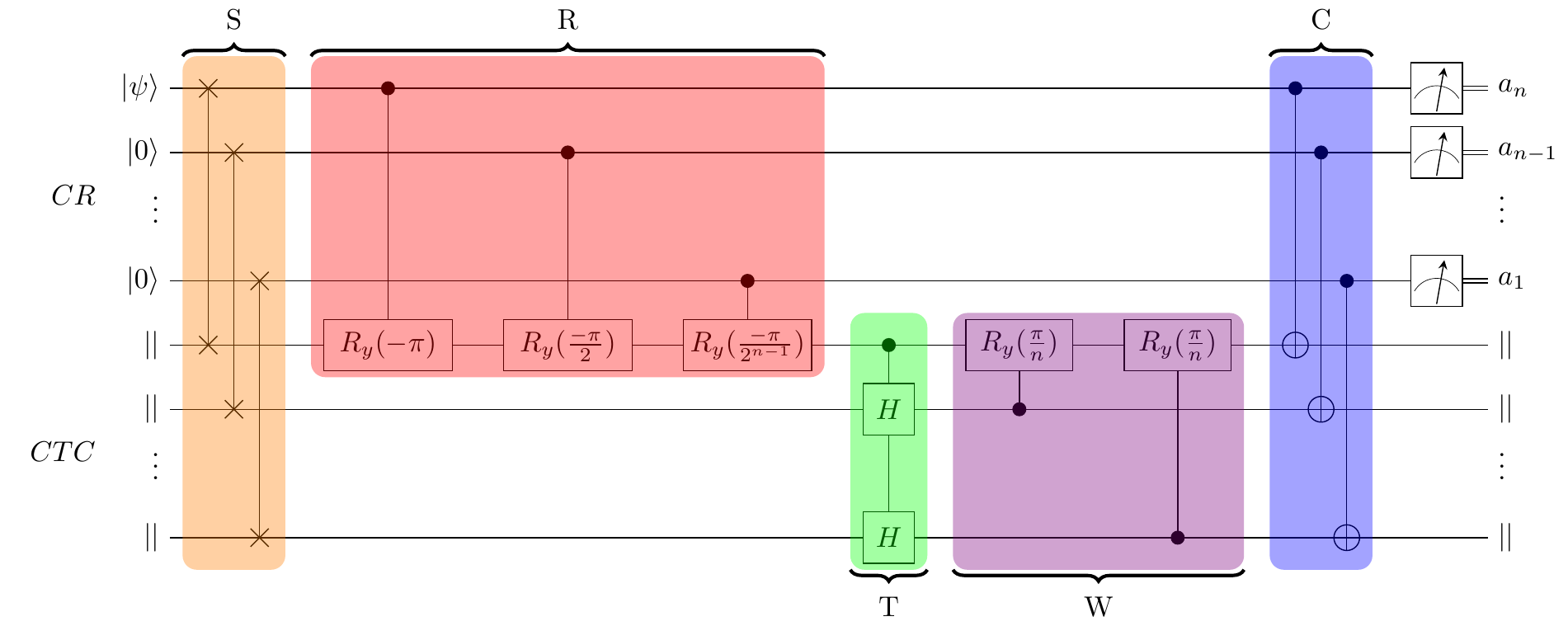}
\caption{\textbf{Retrieval of $n$-bit register from the encoded qubit.} $S$ swaps the CR and CTC registers. $R$ performs a controlled rotation on the encoded qubit. If $R$ does not transform it to $\Ket{0}$, $T$ and $W$ prepare a superposition of all the states in the computational basis of the CTC. $C$ performs a bit-wise controlled not from the CR to the CTC system.}
\label{qdctc_Fig2}
\end{figure*}

We retrieve the encoded information by employing the CTC assisted circuit shown in Fig. \ref{qdctc_Fig2}. We construct the chronology respecting (CR) register and the CTC register containing $n$ qubits each. The first qubit in the CR register $\Ket{\psi}$ is the encoded qubit and the remaining qubits are initialized to state $\Ket{0}$. In the following description, CTC$'$ is used to denote the CTC register excluding the first qubit. The complete interaction unitary is given by $U=CWTRS$ as shown in Fig. \ref{qdctc_Fig2}, 

where
\begin{eqnarray}
S = SWAP(CR \leftrightarrow CTC), \nonumber \\
R = \sum_{j=0}^{2^n-1} \Ket{j}_{CR}\bra{j} \otimes R_y\Big(\frac{-2\pi j}{2^n}\Big) \otimes I_{CTC'}, \nonumber \\
T = I_{CR} \otimes \big(\Ket{0}\bra{0} \otimes I_{CTC'} + \Ket{1}\bra{1} \otimes H^{\otimes (n-1)}\big), \nonumber \\
W = I_{CR} \otimes \sum_{j=0}^{2^{n-1}-1} R_y\Big(\frac{o(j) \pi}{n}\Big) \otimes \Ket{j}_{CTC'}\bra{j}, \nonumber \\
C = \sum_{j=0}^{2^n-1} \Ket{j}_{CR}\bra{j} \otimes \bigg( \sum_{i=0}^{2^n-1} \Ket{i \oplus j}_{CTC}\bra{i} \bigg). \label{qdctc_Eq2}
\end{eqnarray}
Here $\Ket{j}$ represents a state in the computational basis pertaining to the numerical value $j$, $o(j)$ denotes the number of ones in the binary representation of $j$ and $i\oplus j$ denotes bit-wise XOR applied on the binary representations of $i$ and $j$. Let us describe the operation of the circuit by first showing that $\Ket{k}_{CTC}\bra{k}$ is a fixed point of Eq. \eqref{qdctc_Eq1} if $\Ket{\psi} = \Ket{\psi_k}$, where $\Ket{\psi_k}$ is defined in Eq. \eqref{qdctc_psik}. Note that, $\Ket{k}_{CTC} = \Ket{a_n...a_1}_{CTC}$, where $(a_n...a_1)$ is the binary representation of $k$. Let us assume that the state of the encoded qubit is $\Ket{\psi_k}$ and the density matrix of the CTC system exiting the wormhole is $\Ket{k}_{CTC}\bra{k}$. The circuit begins by swapping the CTC register with the CR register (gate $S$). Gate $R$ is similar to the encoding circuit in Fig. \ref{qdctc_Fig1}. The only difference is that the rotation is performed in the opposite direction (clockwise). Thus, the unitary transformation performed on the first qubit of the CTC system is $R_y(-2\pi k / 2^n)$ i.e.,
\begin{equation}\label{qdctc_Eq3}
\Ket{\psi_k} \rightarrow R_y\Big(\frac{-2\pi k}{2^n}\Big)\Ket{\psi_k} = \Ket{0}
\end{equation}
Because the CTC system is now in the state $\Ket{0}^{\otimes n}$, $T$ and $W$ have no influence on the system. Then, the last gate $C$ performs the transformation,
\begin{equation}\label{qdctc_Eq4}
\Ket{k}_{CR}\Ket{0}_{CTC} \rightarrow \Ket{k}_{CR} \Ket{0\oplus k}_{CTC} = \Ket{k}_{CR} \Ket{k}_{CTC}
\end{equation}
By measuring the CR system in the computational basis, we can retrieve the encoded register $(a_n...a_1)$. Furthermore, the density matrix of the CTC system entering the wormhole is $\Ket{k}_{CTC}\bra{k}$, thus ensuring the self-consistency condition. All we need to show now is that this self-consistency solution is unique. Before we proceed, for the sake of convenience, let us introduce the notations:
\begin{eqnarray}
\Ket{\Psi_k} = \Ket{\psi_k}\otimes\Ket{0}^{\otimes n-1}\\
\Ket{\psi(\theta)} = \cos{\frac{\theta}{2}}\Ket{0}+\sin{\frac{\theta}{2}}\Ket{1}
\end{eqnarray}
We can express the total unitary interaction as $U= VS$, where $V$ can be decomposed as
\begin{equation}\label{qdctc_Eq5}
V = \sum_{j=0}^{2^n-1}\Ket{j}_{CR}\bra{j}\otimes V_{j}
\end{equation}
where, $V_{j}$ acts on the CTC. Brun \emph{et al.} \cite{qdctc_BrunPRL2009} describe a set of sufficient conditions for a unique solution of the self-consistency condition, which when applied here has the form
\begin{equation}\label{qdctc_cond}
    \bra{k}V_j\Ket{\Psi_k} \neq 0 \quad \forall \quad k, j
\end{equation}
This implies that if the encoded qubit is $\Ket{\psi_k}$ and the CTC is initialized to $\Ket{j}$, the projection of the final state of the CTC onto $\Ket{k}$ should be non-zero. For $j=k$, we have already seen that this is true. To see that this is true for $j\neq k$, it is convenient to express $R$, $T$ $W$ and $C$ as
\begin{eqnarray}
R = \sum_{j=0}^{2^n-1} \Ket{j}_{CR}\bra{j} \otimes R_j, \nonumber \\
T = I_{CR} \otimes T_{CTC}, \nonumber \\
W = I_{CR} \otimes W_{CTC}, \nonumber \\
C = \sum_{j=0}^{2^n-1} \Ket{j}_{CR}\bra{j} \otimes C_j \label{qdctc_Eq12}
\end{eqnarray}
It is clear that $V_j = C_j W_{CTC} T_{CTC} R_j$. Let us evaluate Eq. \eqref{qdctc_cond} step by step.
\begin{equation}
R_j \Ket{\Psi_k} = \Ket{\psi\Big(\frac{2\pi (k-j)}{2^n}\Big)} \otimes \Ket{0}^{\otimes n-1}
\end{equation}
Let $\theta_{kj} = \frac{2\pi (k-j)}{2^n}$. By applying the operator $T_{CTC}$ on this state, we get,
\begin{equation} \label{qdctc_Eq14}
 \cos{\frac{\theta_{kj}}{2}}\Ket{0}^{\otimes n} + \sin{\frac{\theta_{kj}}{2}}\Ket{1} \otimes \frac{1}{\sqrt{2^{n-1}}}\sum_{i=0}^{2^{n-1}-1} \Ket{i}_{CTC'}
\end{equation}
The first part of the superposition in Eq. \eqref{qdctc_Eq14} remains unchanged under the action of $W_{CTC}$.
\begin{eqnarray}
W_{CTC} \Ket{1}\otimes\Ket{i}_{CTC'} = R_y\Big(\frac{\pi o(i)}{n}\Big) \Ket{\psi(\pi)} \otimes\Ket{i}_{CTC'} \nonumber \\= \Ket{\psi\Big(\pi \Big(1+\frac{o(i)}{n}\Big)\Big)} \otimes\Ket{i}_{CTC'}
\end{eqnarray}

Putting it all together, $W_{CTC} T_{CTC} R_j \Ket{\Psi_k}$
\begin{equation}\label{qdctc_Eq16}
= \cos{\frac{\theta_{kj}}{2}}\Ket{0}^{\otimes n} + \frac{1}{\sqrt{2^{n-1}}} \sin{\frac{\theta_{kj}}{2}} \sum_{i=1}^{2^n-1}\alpha_i \Ket{i}_{CTC}
\end{equation}
where
\begin{eqnarray}\label{qdctc_Eq17}
\alpha_i = \sin{\frac{\pi}{2}\Big(1+\frac{o(i)-1}{n}}\Big) \quad \text{if} \quad i\geq2^{n-1}\nonumber \\
= \cos{\frac{\pi}{2}\Big(1+\frac{o(i)}{n}}\Big) \quad \text{if} \quad i<2^{n-1}
\end{eqnarray}

It can be seen that $\bra{k}C_j = \bra{j \oplus k}$. If $j=k$, then $\theta_{kj} = 0$ and $\bra{k}V_k\ket{\Psi_k} = 1$. If $j\neq k$, then $j \oplus k \neq 0$ and
\begin{equation}\label{qdctc_Eq18}
\bra{k}V_j\ket{\Psi_k} = \frac{1}{\sqrt{2^{n-1}}} \sin{\Big(\frac{\theta_{kj}}{2}\Big)} \alpha_{j\oplus k}
\end{equation}

Since $2\pi > \theta_{kj} > -2\pi$, it is clear that $\sin{(\theta_{jk}/2) = 0}$ only if $j=k$. As for $\alpha_{j\oplus k}$, if $j\oplus k \geq 2^{n-1}$, then $1 \leq o(j\oplus k)  \leq n$. This implies
\begin{equation}\label{qdctc_Eq19}
1 \leq 1+\frac{o(j\oplus k)-1}{n} < 2 \quad \implies \quad \alpha_{j\oplus k} \neq 0
\end{equation}

Similarly, if $j\oplus k < 2^{n-1}$, then $1 \leq o(j\oplus k) \leq n-1$ and thus,
\begin{equation}\label{qdctc_Eq20}
1 < 1+\frac{o(j\oplus k)}{n} < 2 \quad \implies \quad \alpha_{j\oplus k} \neq 0
\end{equation}

We conclude that $\bra{k}V_j\ket{\Psi_k} \neq 0$ for all $j, k$, completing the proof of uniqueness of the solutions to Eq. \eqref{qdctc_Eq1}. We would like to emphasize that the number of 2-qubit gates used in the circuit is $5n-2$. Thus the quantum cost of the circuit in terms of $n$ is $O(n)$.

\section{Simulation of the Circuit \label{qied_Sec4}}
\begin{figure*}[!ht]
\centering
\begin{subfigure}{0.5\linewidth}
\includegraphics[width=\linewidth]{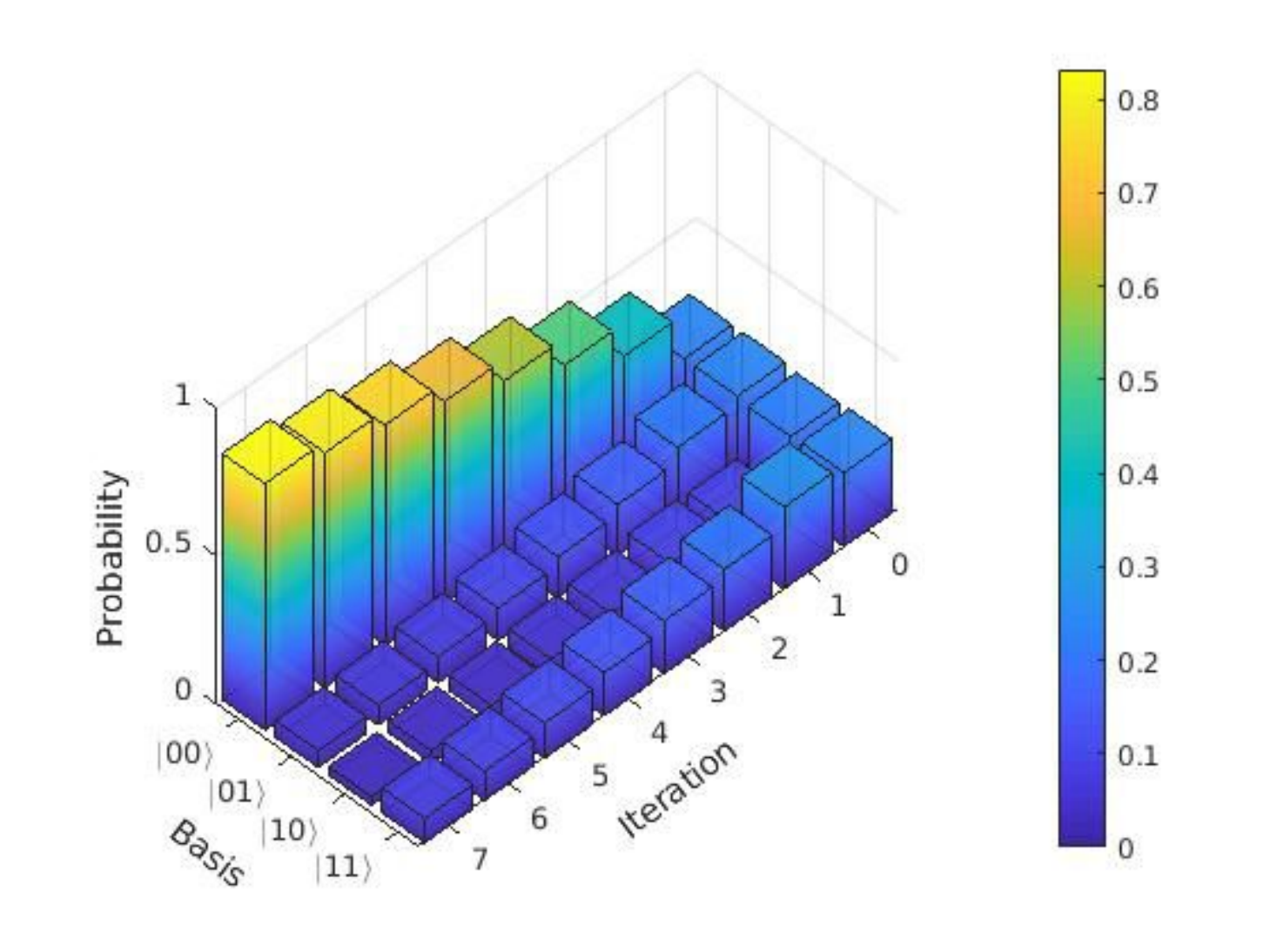} 
\caption{}
\end{subfigure}\hfill
\begin{subfigure}{0.5\linewidth}
\includegraphics[width=\linewidth]{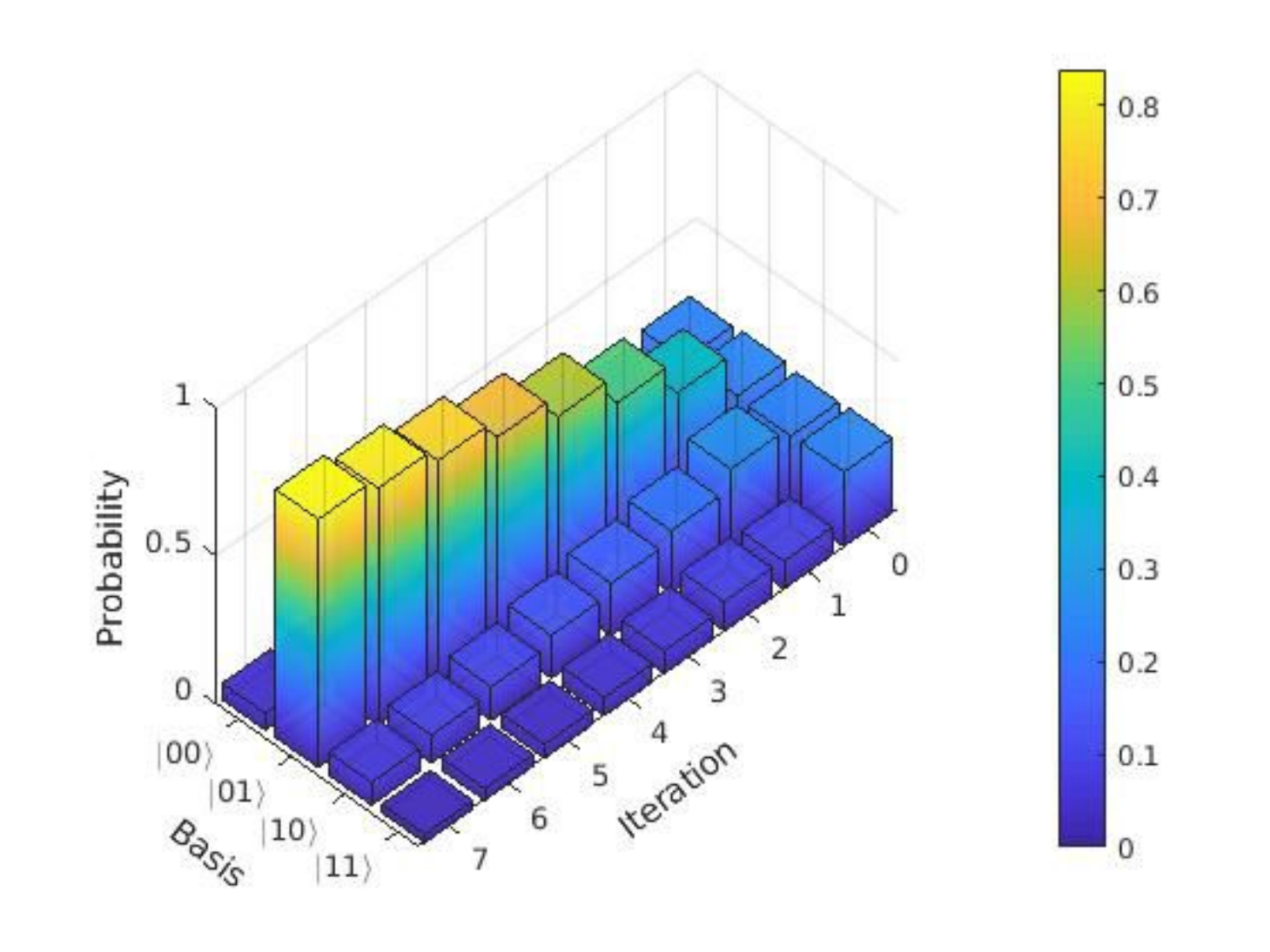} 
\caption{}
\end{subfigure}\hfill
\begin{subfigure}{0.5\linewidth}
\includegraphics[width=\linewidth]{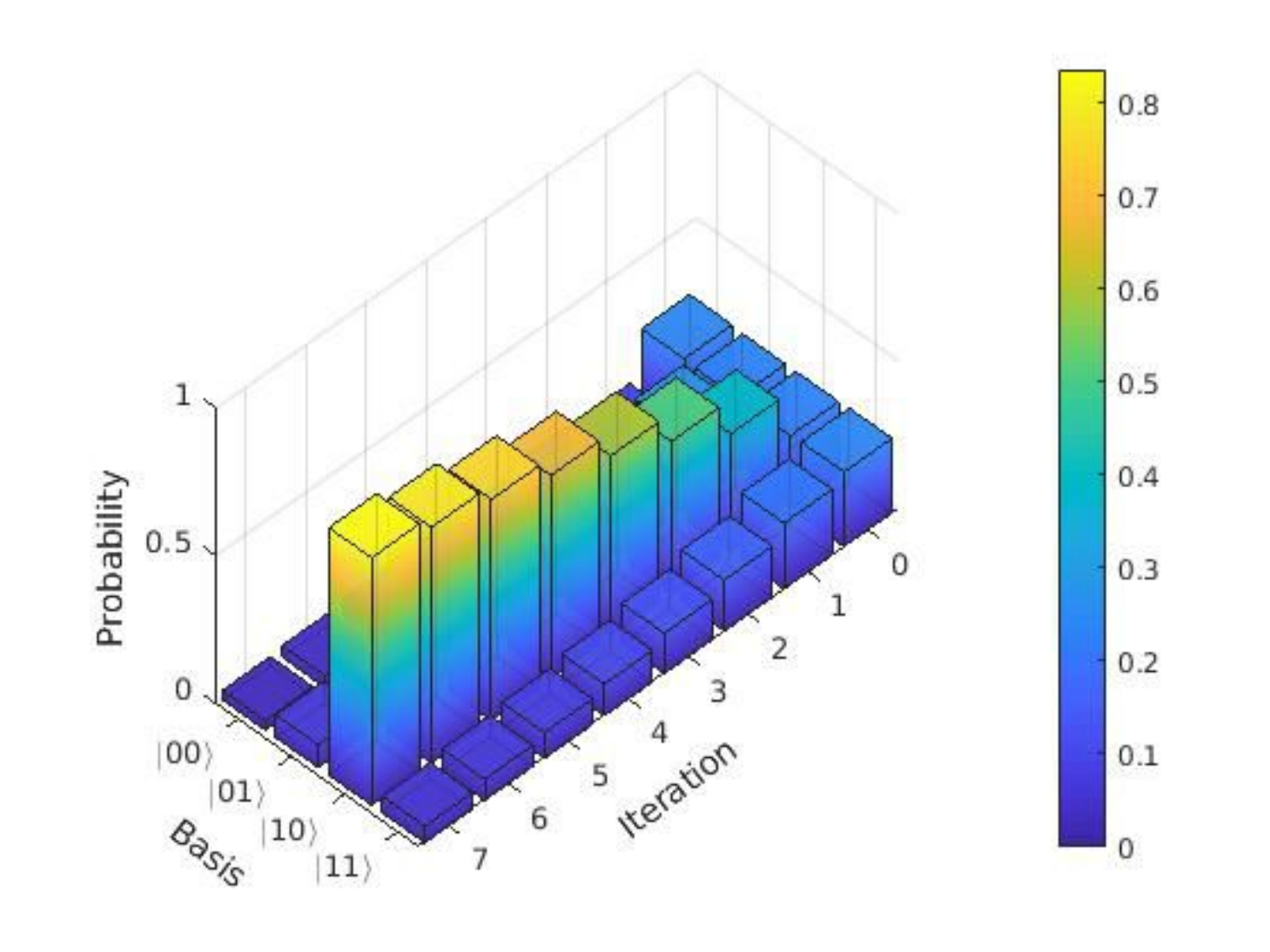} 
\caption{}
\end{subfigure}\hfill
\begin{subfigure}{0.5\linewidth}
\includegraphics[width=\linewidth]{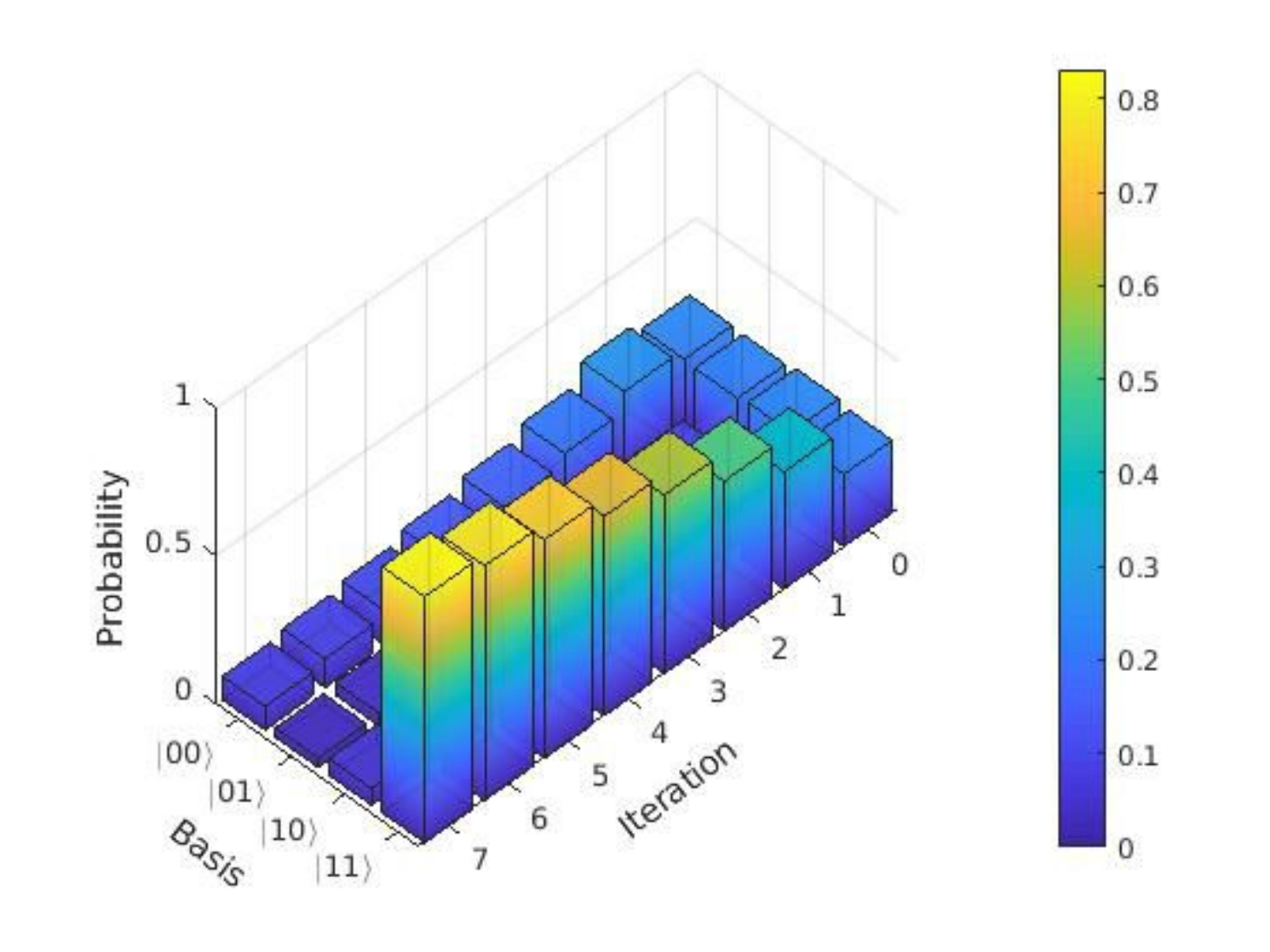} 
\caption{}
\end{subfigure}\hfill
\caption{\textbf{Simulation results of CTC-assisted circuit with 7 iterations.} (a), (b), (c) and (d) correspond to evolution of the CTC system while the encoded states are prepared in $\Ket{0}$, $\Ket{+}$, $\Ket{1}$ and $\Ket{-}$ respectively.}
\label{qdctc_Fig3}
\end{figure*}

Our simulation of the CTC-assisted circuit is similar to the procedure given by Brun and Wilde \cite{qdctc_BrunFoundPhys2017}. Consider the map $\mathcal{N}_{U, \rho}$, 
where
\begin{equation}\label{qdctc_Eq6}
\mathcal{N}_{U, \rho}(\omega_{CTC}) = Tr_{CR}(U(\rho_{CR}\otimes\omega_{CTC})U^\dagger)
\end{equation}
The consistency condition states that nature initializes the CTC system such that its density matrix is a fixed point of $\mathcal{N}_{U, \rho}$; i.e., a state $\sigma_{CTC}$ such that
\begin{equation}\label{qdctc_Eq7}
\mathcal{N}_{U, \rho}(\sigma_{CTC}) = \sigma_{CTC}
\end{equation}
For our circuit, we see that for any initial state, say $\omega_{CTC}$,
\begin{equation}\label{qdctc_Eq8}
\lim_{N\rightarrow\infty} \mathcal{N}^N_{U, \rho}(\omega_{CTC}) = \sigma_{CTC}
\end{equation}

\begin{figure}[]
\centering
\includegraphics[width=\linewidth]{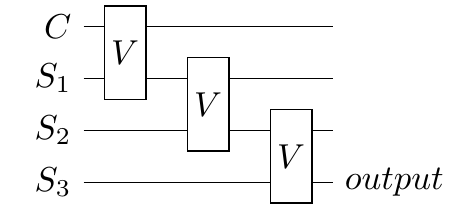}
\caption{\textbf{Simulation of the CTC-assisted circuit.} Here $C$ is assigned to an arbitrary state. $S_1$, $S_2$ and $S_3$ are prepared in the state $\rho_{CR}$. This circuit depicts three iterations.}
\label{qdctc_Fig4}
\end{figure}

\begin{figure*}[!ht]
\centering
\includegraphics[width=\linewidth]{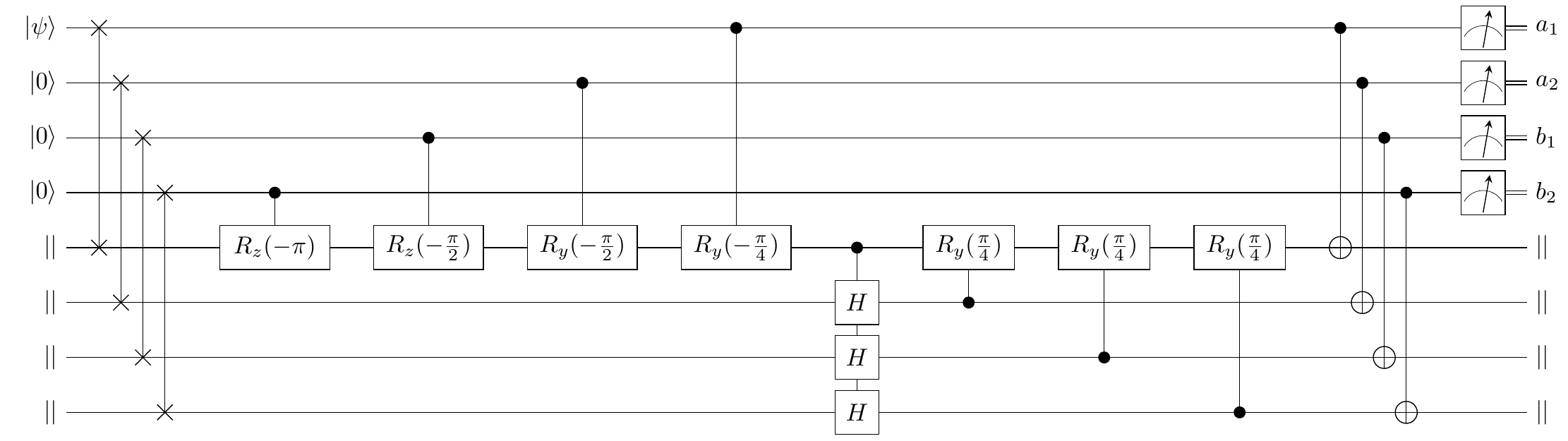}
\caption{\textbf{CTC-assisted circuit to clone a single qubit state.} $\Ket{\psi}$ is the state to be cloned. The number of qubits in the CTC representing the azimuthal and polar angles are 2 each. Measurement on the first and the last two qubits gives the information about the polar angle and the azimuthal angle of $\Ket{\psi}$ respectively.}
\label{qdctc_Fig5}
\end{figure*}
Implying that consequent applications of the quantum channel $\mathcal{N}_{U, \rho}$ on $\omega_{CTC}$ cause it to converge to $\sigma_{CTC}$. Inspired by this and also by noting that $U = VS$, where $V$ represents the gates of the interaction unitary excluding the swap gate, we constructed the circuit given in Fig. \ref{qdctc_Fig4} to simulate the CTC-assisted circuit. Here $C$ is initialized to any arbitrary state to represent the CTC system. $S_1$, $S_2$ and $S_3$ are individually initialized to the state $\rho_{CR}$. Note that multiple copies of the CR system are required for the simulation. We have constructed the simluation circuit of a 2-bit decoding circuit for different number of iterations and ran it on the \textit{IBM QASM Simulator}. The initial state of $C$ was prepared as a uniform superposition of 2-qubit computational basis states by applying a Hadamard on each qubit. For the case of 2-qubits, the representative states are $\Ket{0}$, $\Ket{1}$, $\Ket{+} = \frac{1}{\sqrt{2}}(\Ket{0}+\Ket{1})$ and $\Ket{-} = \frac{1}{\sqrt{2}}(\Ket{0}-\Ket{1})$. The results of the simulations are shown in Fig. \ref{qdctc_Fig3}. It is observed that as the iterations increase, the output state converges to the required fixed point state.

\section{Cloning Circuit}

Based on the ideas developed in the previous sections, we can construct a CTC-assisted circuit to clone a qubit. Although, the cloning is imperfect, the fidelity of cloning converges to 1 as the number of ancillary qubits increases. For the decoding circuit, the distinguishable states are present on the XZ plane of the Bloch sphere. The measurements yield the polar angle of the state (when represented on a Bloch sphere), which lies in the range $[0, 2\pi)$. This is because we perform rotations on the encoded qubit only along the Y-axis.

The aim of our cloning scheme is to figure out the polar and the azimuthal angles of the state to be cloned. Once, they are known we can reconstruct the state onto a target qubit. We require some qubits in the CTC system to denote the polar angle and some to denote the azimuthal angle. Let $n$ and $m$ be the number of qubits representing the polar and azimuthal angles respectively. We construct a map such that the polar angle lies in the range $[0, \pi)$, while the azimuthal angle lies in the range $[0, 2\pi)$. In the circuit, the azimuthal qubits control the rotation along the Z-axis and the polar qubits control the rotation along the Y-axis. The Z-axis rotation is followed by the Y-axis rotation. The remainder of the circuit is the same as the decoding circuit. Fig. \ref{qdctc_Fig5} shows the circuit for $n=2$ and $m=2$. The measurements $(a_2, a_1)$ and $(b_2, b_1)$ correspond to the polar angle and the azimuthal angle respectively. Note that the state $\Ket{1}$ does not belong to the set of perfectly distinguishable states and thus cannot be perfectly cloned. Also, the state $\Ket{0}$ can be constructed with a polar angle of 0 and any azimuthal angle, and thus for this case, the consistency condition yields multiple fixed points for the CTC system. However, all the fixed points construct $\Ket{0}$ and thus the fidelity of cloning $\Ket{0}$ is 1.

\begin{figure*}[!ht]
\centering
\begin{subfigure}{0.5\linewidth}
\includegraphics[width=\linewidth, trim={200, 10, 200, 10}, clip]{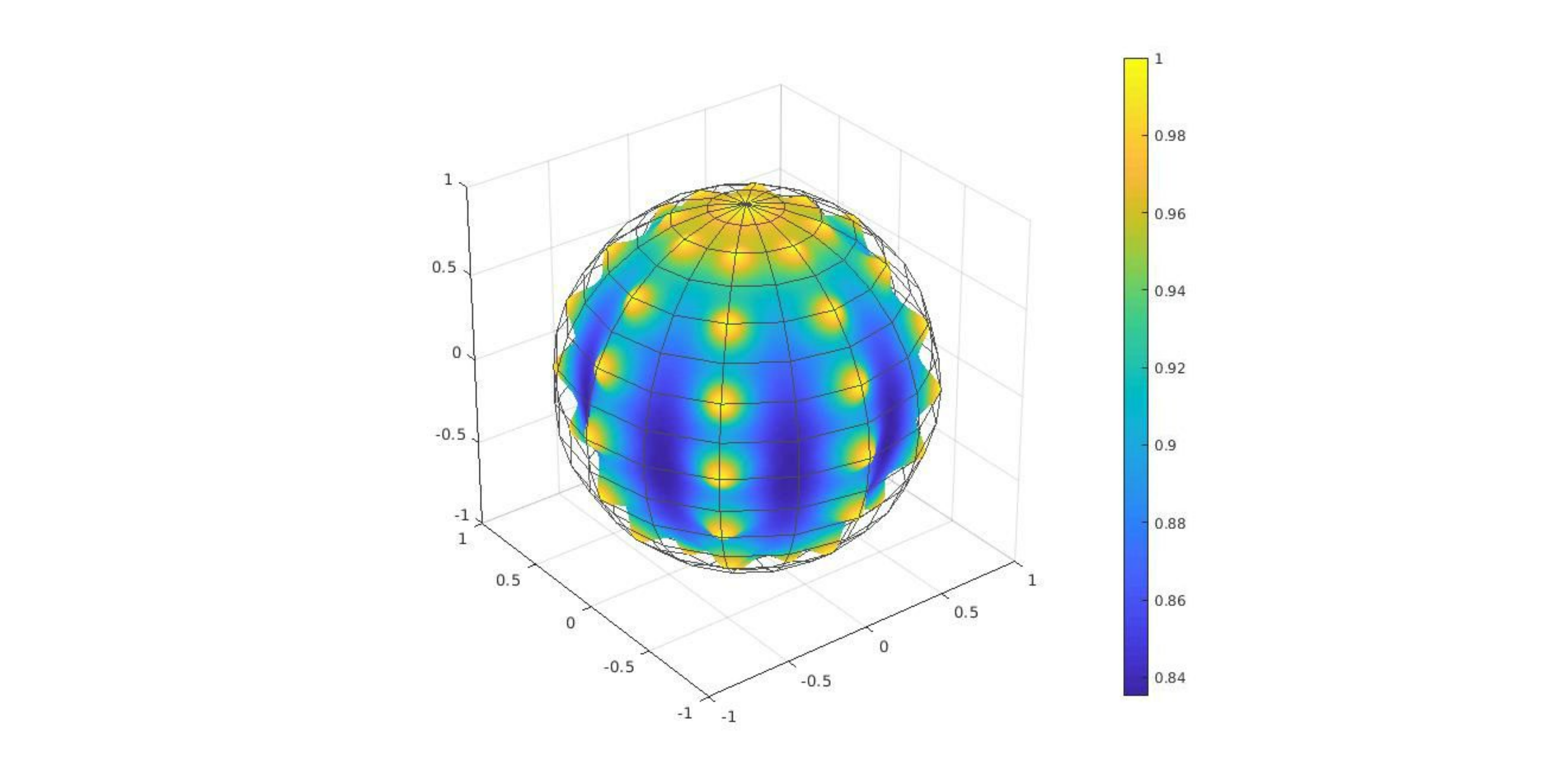} 
\caption{}
\label{qdctc_Fig6a}
\end{subfigure}\hfill
\begin{subfigure}{0.5\linewidth}
\includegraphics[width=\linewidth, trim={200, 10, 200, 10}, clip]{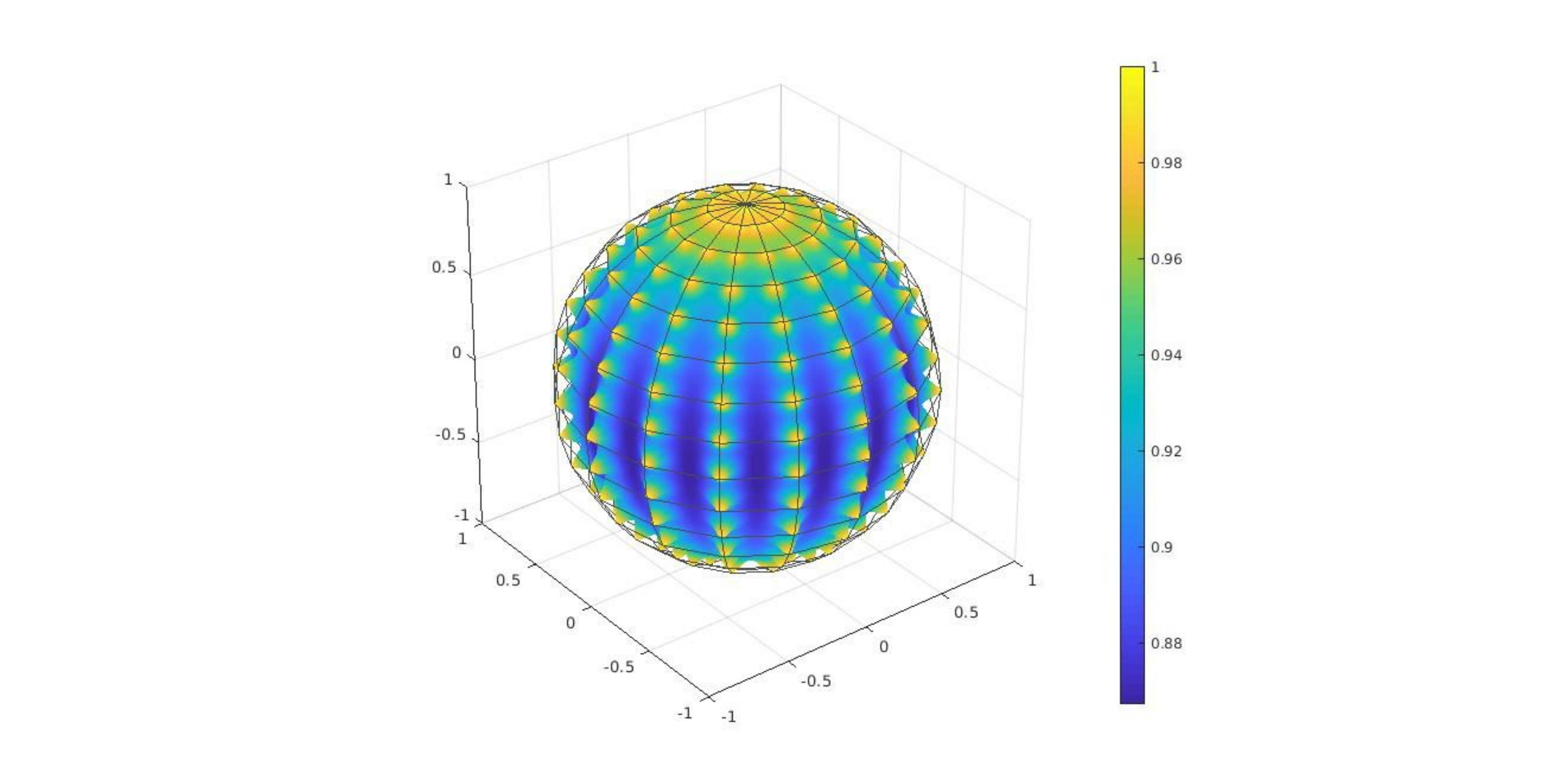} 
\caption{}
\label{qdctc_Fig6b}
\end{subfigure}\hfill
\caption{\textbf{Surface plot of the fidelity of cloning on the Bloch sphere.} (a) and (b) correspond to the circuit with $n=m=3$ and $n=m=4$ respectively. The points with the fidelity 1 (yellow) form the set of states which can be perfectly distinguishable, and hence can be perfectly cloned. It can be observed that as $n$ and $m$ increase, the number of perfectly distinguishable points increases.}
\label{qdctc_Fig6}
\end{figure*}

Once the measurements are done, the reconstruction yields only the states that are perfectly distinguishable. Thus, the states that are not perfectly distinguishable cannot be cloned with a fidelity of 1. We evaluate the fidelity of cloning such states by first evaluating the fixed point density matrix of the CTC system for each state followed by reconstructing the density matrix of the target qubit based on this fixed point. Finally we compare the density matrix of the target qubit with the pure state of the original qubit by evaluating the fidelity between them. Fig. \ref{qdctc_Fig6} shows the fidelity of cloning for the states on the Bloch sphere. The points that have a fidelity of 1 form the set of perfectly distinguishable states. Fig. \ref{qdctc_Fig6a} corresponds to a circuit with $n=3$ and $m=3$ and Fig. \ref{qdctc_Fig6b} corresponds to a circuit with $n=4$ and $m=4$. The quantum cost of this circuit in terms of $n$ and $m$ is $O(n+m)$.

\section{Conclusion}
To conclude, we have proposed here a new scheme for storing an n-bit classical register in a single qubit and retrieving the information faithfully in the presence of a D-CTC. The scheme clearly violates the Holevo bound. Furthermore, the quantum cost of the scheme is of the order $O(n)$, while the existing schemes have the order of $O(2^n)$. A simulation of 7 iterations for retrieving the stored information has been demonstrated. We have also shown that the scheme can be modified to clone a qubit. Although the cloning process is imperfect, it has been observed that as the number of qubits in the CTC system representing the azimuthal and polar angles of the qubit (to be cloned) increases, the average fidelity of cloning tends to 1. The proposed scheme for cloning has an advantage that one can easily increase the number of ancillary qubits as the quantum cost is linear in terms of the number of ancillary qubits, in contrast to existing schemes, which have an exponential quantum cost.

\section*{Acknowledgements}H.R.N. acknowledges the hospitality of Indian Institute of Science Education and Research Kolkata during the project work. B.K.B. acknowledges the financial support of IISER Kolkata. We acknowledge the support of IBM Quantum Experience for using the quantum processors. The views expressed are those of the authors and do not reflect the official position of IBM or the IBM quantum experience team.

\end{document}